Copyright Notices

# Liquid-Vapor Phase Equilibrium in Molten Aluminum Chloride (AlCl$_3$) Enabled by Machine Learning Interatomic Potentials

*Rajni Chahal,[1*] Luke D. Gibson,[2] Santanu Roy,[1] Vyacheslav S. Bryantsev[1*]*

[1]Chemical Science Division, Oak Ridge National Laboratory, Oak Ridge, TN-37830, United States

[2]Computational Sciences and Engineering Division, Oak Ridge National Laboratory, Oak Ridge, TN 37831, USA

chahalr@ornl.gov, bryantsevv@ornl.gov

## ABSTRACT

Molten salts are promising candidates in numerous clean energy applications, where knowledge of thermophysical properties and vapor pressure across their operating temperature ranges is critical for safe operations. Due to challenges in evaluating these properties using experimental methods, fast and scalable molecular simulations are essential to complement the experimental data. In this study, we developed machine learning interatomic potentials (MLIP) to study the AlCl$_3$ molten salt across varied thermodynamic conditions (T=473–613 K and P=2.7-23.4 bar), which allowed us to predict temperature-surface tension correlations and liquid-vapor phase diagram from direct simulations of two-phase coexistence in this molten salt. Two MLIP architectures, a Kernel-based potential and neural network interatomic potential (NNIP), were considered to benchmark their performance for AlCl$_3$ molten salt using experimental structure and density values. The NNIP potential employed in two-phase equilibrium simulations yields the critical temperature and critical density of AlCl$_3$ that are within 10 K (~3%) and 0.03 g/cm$^3$ (~7%) of the reported experimental values. An accurate correlation between temperature and viscosities is obtained as well. In doing so, we report that the inclusion of low-density configurations in their training is critical to more accurately represent the AlCl$_3$ system across a wide phase-space. The MLIP trained using PBE-D3 functional in the ab initio molecular dynamics (AIMD) simulations (120 atoms) also showed close agreement with experimentally-determined molten salt structure comprising Al$_2$Cl$_6$ dimers, as validated using Raman spectra and neutron structure factor. The PBE-D3 as well as its trained MLIP showed better liquid density and temperature correlation for AlCl$_3$ system when compared to several other density functionals explored in this work. Overall, the demonstrated approach to predict vapor pressure and temperature correlations in this study can be employed to screen nuclear reactors-relevant compositions, helping to mitigate safety concerns.

**INTRODUCTION**

Molten salts due to their favorable chemical, physical, and thermal properties lend themselves to numerous high-temperature applications, such as nuclear reactor coolants/fuels [1], [2], concentrated solar plants [3], and high-temperature batteries. Under extreme conditions, the increase in the temperature can cause the vapor pressures of molten salts to build up, leading to serious safety concerns [2]. $AlCl_3$ is an example of a volatile molten salt system, where temperature changes can significantly impact the vapor pressure [4], [5]. For this reason, $AlCl_3$ was selected in the present study as a specific case to demonstrate the feasibility of our approach to compute the liquid-vapor phase equilibrium from molecular simulations using machine learning interatomic potentials (MLIPs). Typically, the increased difficulty in collecting experimental data at higher temperatures [6], [7] lead to substantial uncertainties [8], [9], [10] in vapor pressure measurements. For example, these uncertainties contribute to an error of at least ±100 K for the boiling point of $2LiF-BeF_2$ in the MSTDB-TC database [11].

The commonly utilized computational methods to estimate the vapor pressure from molecular simulations include direct coexistence simulations [12], Gibbs ensemble Monte Carlo [13], and those computing the difference in the chemical potentials in two phases [14]. While a significant amount of literature exist on the application of these methods to other materials [12], [15], [16], [17], [18], [19], only a few studies were conducted for molten salts, using mostly a semi-empirical Computer Coupling of Phase Diagrams and Thermochemistry (CALPHAD) approach [20], [21]. Due to the requirement of large system sizes and long simulation times, density functional theory (DFT)-based ab initio molecular dynamics (AIMD) simulations are impractical to reliably predict the vapor pressure (along with other properties of interest) with an acceptable statistical error. Furthermore, the choice of DFT methods used to study a given material system is known to significantly impact both its structure and properties, leading to varying levels of accuracy [22], [23], [24]. In our previous work, we have used X-ray and neutron scattering along with Raman spectroscopy to identify a suitable DFT method to predict the structure of molten salts [25], [26], [27]. Such a validation for salt structure is critical for molten salt systems due to the existence of a strong structure-property relationship, necessitating an accurate prediction of their structure to accurately predict their properties in the desired temperature ranges [6], [27]. Consequently, it is critical when modeling pure $AlCl_3$ that the chosen DFT method accurately predicts the stability of $Al_2Cl_6$ dimers, which is confirmed by experimental neutron structure factors and Raman spectra measurements [28], [29], [30].

Previously, the Reverse Monte Carlo (RMC) method failed to reproduce the dimeric structure of pure $AlCl_3$ [31] because the total structure function is insensitive to Al-Al correlations [32]. Using a polarizable ion model (PIM) [33] that was developed for single-component systems, classical molecular dynamics simulations correctly predicted stable $Al_2Cl_6$ dimers in pure $AlCl_3$ and reproduced the X-ray and neutron-weighted structure factors [34], [35], [36]. However, a more recent PIM model [37] trained on 50% mixture of $AlCl_3$ with monovalent metal ions yielded a significant portion of larger clusters [26]. Such disagreement in the structure prediction of molten salts involving multivalent cation species has been previously shown to have a significant influence on the predicted transport and thermophysical properties of the salts [6], [38], [39], [40]. Therefore, to accurately predict essential thermodynamic quantities directly from the atomistic simulations, it is imperative to obtain the correct salt structure and speciation [26]. Notably, DFT-trained MLIPs—which dramatically increase the accessible time- and length-scales of molecular dynamics simulations [41]—have been shown to accurately predict the structure of molten salts, as evidenced by the excellent agreement with several experimental measurements (e.g., X-ray, neutron, and Raman spectra [6], [7], transport and thermophysical properties [42], [43], [44], [45]).

Motivated by the success of MLIPs in these studies, we aim to address the aforementioned limitations in predicting pure $AlCl_3$ structure, densities at the liquid-vapor equilibrium, viscosity, and surface tension by developing MLIPs from the AIMD data. Overall, this study establishes and demonstrates the utility of machine learning potentials in assessing liquid-vapor phase equilibria, provided that low-density cluster configurations are included in the training. The developed methodology can be extended to study the phase equilibria in other complex binary and ternary molten salt systems, such as the melts commonly found in molten salts nuclear reactors where harsh conditions (e.g., radioactive environment and high temperatures) prevent reliable measurements [6], [7], [25], [46].

## COMPUTATIONAL METHODS

In this section, we discuss the computational methods used to determine the AlCl$_3$ structure and properties. First, we explored different DFT methods to identify the most accurate exchange-correlation functional for the AlCl$_3$ system that will be employed for MLIPs training. After identifying the optimal DFT method, we developed MLIPs for the AlCl$_3$ system using two different architectures. The first architecture we employed was the kernel-based MLIP implemented in VASP, known as a machine learning force field (MLFF), which was trained using the integrated on-the-fly active learning scheme [47]. The other architecture was a neural network interatomic potential (NNIP), which was trained using the DeePMD-kit package [48], [49]. The details of the subsequent molecular dynamics simulations using trained MLFF and NNIP potentials are provided next.

### Identifying Optimum Exchange-Correlation Functionals

To identify the best performing exchange-correlation (XC) functional for the AlCl$_3$ system, CP2K [50] was used to run AIMD simulations across a wide span of functional classes listed in **Table 1**. Van der Waals (vdW) interactions were treated using various levels of theory. Grimme's

D3 method [51] was used in conjunction with PBE [52], [53], [54] and SCAN[55], which applies a post hoc empirical correction to the electronic energy. Non-local vdW density functionals [56], [57](SCAN-RVV10 [58], optB86b-vdW [59], and optB88-vdW [60]) were also considered due to their more sophisticated inclusion of dispersion interactions directly into the SCF calculations. Predicted Raman spectra and structure factors were computed from AIMD simulations in the NVT ensemble, which contained 30 $AlCl_3$ pairs (15 $Al_2Cl_6$ dimers) in a 17.62 x 17.62 x 17.62 $Å^3$ cell to match the experimental density at 498 K (1.21 $g/cm^3$). The Nosé-Hoover thermostat [61] with a 1 ps time constant was used to maintain the system temperature for at least 100 ps with all employed DFT methods. Constant pressure [62] simulations (NPT ensemble) were then continued for at least another 100 ps at 498 K and 4.33 bar, which corresponds to the vapor pressure at the experimental density [5] (given by **Equation 1**). Based on the superior performance of PBE-D3, we employed this functional when generating the AIMD data for training both NNIP (using CP2K data) and MLFF (using VASP data) models.

**Table 1**. Exchange-correlation (XC) functionals and their corresponding types of van der Waals interactions and DFT functional class considered in CP2K AIMD simulations.

| XC Functional | Inclusion of vdW Interactions | Functional Class |
| --- | --- | --- |
| **PBE-D3** | Empirical | GGA |
| **SCAN** | - | meta-GGA |
| **SCAN-D3** | Empirical | meta-GGA |
| **SCAN-RVV10** | Non-local | vdW-DF (meta-GGA) |
| **optB86b-vdW** | Non-local | vdW-DF (GGA) |
| **optB88-vdW** | Non-local | vdW-DF (GGA) |

**VASP Machine-Learned Force Field (MLFF) Development**

*Data generation: Ab-Initio Molecular Dynamics Simulations using VASP*

For on-the-fly training of the MLFF [47], AIMD simulations using D3 dispersion [51] were performed using VASP version 6.3 [63], [64], [65], [66], [67]. The valence electrons were expanded with plane waves and the core-valence interactions were described using the projector augmented wave (PAW) method [68], [69]. Including semi-core $2p$ states as valence states for aluminum did not noticeably change the results; therefore, the aluminum pseudopotential used in VASP only included $3s^2 3p^1$ valence states. A plane wave kinetic energy cutoff of 600 eV was used during training to permit flexible cell dimensions. The electronic energy convergence threshold was set to $10^{-5}$ eV for self-consistent field (SCF) cycles. Charge density mixing followed the Pulay mixing scheme [70] during SCF calculations. In all cases, a 1 x 1 x 1 k-point mesh grid was used.

*MLFF Training*

The MLFF interatomic potential was generated using the on-the-fly training algorithm natively implemented in VASP version 6.3. On-the-fly training involves a molecular dynamics (MD) simulation that dynamically trains the MLFF on DFT data for frames that have estimated errors that exceed a threshold. The MLFF in VASP closely follows the Gaussian Approximation Potential (GAP) formalism [71] and is a kernel-based method. A detailed explanation of the method implementation in VASP is provided in Ref. [47]. Briefly, the method uses radial and angular descriptors to characterize the local environment around an atom within a cutoff radius. These descriptors are then compared to reference configurations in the training set to estimate the energy and forces for that atom. In this formalism, the major hyperparameter that controls the accuracy of the MLFF is the limit on the number of local reference configurations (LRCs) in its database. However, the computational expense to run the MLFF also increases as more LRCs are included. Each atom type has its own set of LRCs and, thus, the speed of the MLFF also drops off as more atom types are included. In this work, we limited the maximum number of LRCs to 4,000 and used an 8 Å cutoff radius for both radial and angular descriptors. Using the on-the-fly training workflow, we performed isobaric-isothermal MD (NPT) for a system of 15 $Al_2Cl_6$ dimers with a 1 fs time step and a temperature that was linearly ramped from 500 K to 800 K over 50 ps. The Parrinello-Rahman method [72], [73] was used to maintain the temperature and pressure during MD with Langevin friction coefficients of 5 $ps^{-1}$ for both quantities. Due to the nature of this active learning approach, a vast majority of the frames were evaluated with the generated MLFF instead of the expensive electronic structure calculations. Additional NVT simulations of systems with densities of 0.8 and 0.6 $g/cm^3$ were included to bolster the model's ability to perform in low density regions. The final MLFF contained 1,555 frames from the training MD trajectories and the number of local reference configurations for Al and Cl were 2,768 and 2,698, respectively. The final root mean squared errors (RMSE) for training of MLFF were 0.84 meV/atom, 38.3 meV/Å, and 0.14 kbar for energy, forces, and stresses, respectively. For the test dataset, the energy, forces, and stresses RMSE errors were 0.96 meV/atom, 33.5 meV/Å, and 0.12 kbar, respectively. These low error rates demonstrate the MLFF's ability to closely mimic PBE-D3 accuracy.

*MLFF-based Molecular Dynamics (MLMD) Simulations*

The trained MLFF potential was used in VASP to equilibrate a periodic system of 100 $Al_2Cl_6$ dimers randomly packed in a simulation cell corresponding to the experimental density at a given temperature [74]. To obtain equilibrium densities of the liquid phase, constant pressure simulations were performed at 473 K, 498, K, 508 K, 543 K, 598 K, and 613 K. At each temperature, the system pressure was set to the corresponding vapor pressure obtained from the relationship established by Viola et al. [5] (given by **Equation 1**), where $P$ is the vapor pressure in Torr and $T$ is the temperature in K.

$$\log_{10} P = -\frac{1956.06}{T} + 7.43530 \tag{1}$$

Bulk $Al_2Cl_6$ systems were equilibrated for at least 2 nanosecond (ns) using the Parrinello-Rahman method to maintain appropriate temperatures and pressures with Langevin friction coefficients of 5 $ps^{-1}$ [72], [73]. After equilibration, 1 ns trajectories were generated for all temperature and pressure conditions to study the salt structure and melt densities.

**DeePMD Neural Network Interatomic Potential (NNIP) Development**

*Data generation: Ab-Initio Molecular Dynamics Simulations using CP2K*

To generate NNIP training data, Born-Oppenheimer AIMD simulations were performed using the QUICKSTEP module of the CP2K package [50], the double-zeta MOLOPT basis sets (DZVP-MOLOPT) with a density cutoff of 600 Ry, and Goedecker-Teter-Hutter (GTH) pseudo-potential. CP2K simulations with Perdew, Becke, and Ernzerhof (PBE) functional and Grimme's D3 correction [50], [51] were used to generate all the training data for NNIP potentials. The canonical ensemble (NVT) using a Nosé-Hoover chains thermostat [75] with 100 fs time constant was employed while maintaining the periodic boundary conditions. Several AIMD simulations were performed using 1 fs timestep for 16-80 ps, specific details are provided in the next section.

*NNIP Training*

To develop NNIP models, we employed the Deep-Pot-Smooth Edition (DeepPot-SE) potential [49] contained inside DeePMD-kit package (version 2.1.1) [48]. The DeepPot-SE has been previously demonstrated to fit a continuously differentiable potential energy surface for multicomponent molten salts [6], [7], [40] and other complex systems [41]. To train the initial NNIP (referred as NNIP-1) for pure $AlCl_3$, a total of 55,355 configurations of 15 $Al_2Cl_6$ dimers were included in the dataset, which contained a range of densities around the equilibrium density at 498 K. As previously pointed in [6], [42], it is crucial to include bulk configurations from the same system with varying densities during training to ensure that the resulting NNIP produces stable densities in NPT simulations. To achieve this, we included bulk configurations with either expanded or compressed volumes (relative to experimental densities at 498 K) in the following amounts: 16,074 (5% compressed), 16,125 (5% expanded), 19,123 (10% compressed), 16,513 (10% expanded), and 71,000 (25% expanded). The second NNIP (NNIP-2) was developed by including 454,839 additional configurations beyond the NNIP-1 dataset. Specifically, bulk configurations with further expanded or compressed volumes were included in the following amounts: 51,800 (20% compressed), 57,620 (30% compressed), 38,380 (40% expanded), and 81,333 (70% expanded). Lower density configurations were also included to enhance the NNIP's reliability in the vapor region. For this purpose, periodic systems containing 2, 4, and 7 $Al_2Cl_6$ dimers were simulated. Gas phase configurations were selected in the following amounts: 28,200 (2 dimers at $\rho$ = 0.16 g/cm$^3$ and 498 K), 80,000 (4 dimers at $\rho$ = 0.32 g/cm$^3$ and 621 K), 80,000 (7 dimers at $\rho$ = 0.57 g/cm$^3$ and 628 K), and 37,506 (7 dimers at $\rho$ = 0.28 g/cm$^3$ and 628 K). For both NNIP-1 and NNIP-2 models, the datasets were shuffled and split, with 80% and 20% representing the training and validation sets, respectively. Overall, the training data was selected based on training-validation-augmentation procedures previously used to develop robust NNIPs [6].

The DeepPot-SE model learns a mapping between the local environment of each atom within a cutoff (here, 8 Å) and a per-atom energy, where the sum of atomic energies corresponds to the reference DFT energy [48], [49]. Thereafter, the gradients of the NNIP-predicted energies are used to compute the atomic forces. Here, both the reference energies and forces are included to evaluate the loss function which is minimized during training of the model. In the model, the smooth and hard cutoff radii are chosen to be 2 and 8 Å, respectively. The embedding and fitting network sizes are taken as {25,50,100} and {240,240,240}, respectively. The tunable prefactors in the loss function are 0.002, 1000, 1, 1 for $p_e^{start}$, $p_f^{start}$, $p_e^{limit}$, and $p_f^{limit}$, respectively. These hyperparameters have previously demonstrated success in accurately predicting the structure and transport properties of multicomponent systems [6]. The weights and biases of the neural network were initialized with random values using a seed value of 1. Both trainings were continued for 800,000 steps. After training, NNIP-1 yielded average energy and force errors of 0.52 meV/atom and 29.6 meV/Å, respectively, on the validation set. For NNIP-2, the average energy and force errors on the validation set increased only slightly to 0.55 meV/atom and 30.9 meV/Å, respectively, demonstrating a good fit to more complex DFT potential energy surfaces. The energy errors on the training and validation sets for all systems corresponding to the marked datapoints in **Figure S1** are plotted in **Figure S2(a-b)**. In addition, **Figure S2(c-f)** provides a comparison of the energy errors for dimer configurations between the two NNIPs.

Finally, to benchmark the transferability of the trained MLFF and both NNIP potentials to untrained structures, we constructed a test set consisting of 300 frames. The test set comprised 100 frames each taken uniformly from three NVT MD simulations with densities of 0.79, 1.10, and 1.27 g/cm$^3$. As shown in **Figure S3**, MLFF, NNIP-1, and NNIP-2 potentials are able to closely reproduce the DFT forces with the test error of 25.1, 24.1, and 25.9 meV/Å, respectively.

***NNIP-based Molecular Dynamics (NNMD) Simulations***

*Bulk Liquid Densities Calculations*

The trained potentials (NNIP-1 and NNIP-2) were used in LAMMPS [76] *via* the interface with DeePMD-kit [76]. Here, the NNIPs were first used to equilibrate a periodic simulation system containing 120 $Al_2Cl_6$ dimers in order to obtain the relaxed densities for the liquid phase at 473, 498, 508, 543, 598, and 613 K with the system pressures set to the corresponding vapor pressures from **Equation 1**. During the NNMD runs, bulk $Al_2Cl_6$ systems were relaxed for at least 5 ns at specified temperature and pressure values using Nosé-Hoover thermostat/barostat (NPT) [61], [77] with a 100 fs time constant for both thermal and pressure baths. After this step, 3-8 ns trajectories were generated to study the salt structure as well as for obtaining molten salts densities.

*Liquid and Vapor Densities from Two-Phase Simulations*

In order to prepare a vapor-liquid interfacial system, a liquid box of 960 $Al_2Cl_6$ dimers at the average equilibrated density (obtained from the NPT simulation) was expanded along the z-dimension to produce a liquid slab in the x- and y-dimensions. We note that the z-dimension was selected such that the average simulation box density was within the coexistence region on the phase diagram (**Figure S4**) for each temperature.

First, NPT simulations were performed with 1 fs time step to converge the bulk liquid phase density. Then, the z-dimension was extended to lower the average density of the system. The two-phase system was then equilibrated in the NVT ensemble for 8-18 ns [77], during which the density profiles along the z-direction were obtained every 100 fs. To account for the slab's movement along the z-direction during equilibration, the entire system was repositioned so the center of the liquid slab was at half the z-dimension of the simulation box (i.e., z = 0 in **Figure S5**) for each density profile. This was necessary to avoid error accumulation while obtaining the averaged density profile from the equilibrated trajectory.

*Viscosity Calculation*

Green-Kubo method [78] was used to evaluate the viscosity of $AlCl_3$ at 473 and 498 K by calculating the time integral of the shear stress autocorrelation function (ACF) according to **Equation 2**.

$$\eta = \frac{V}{k_B T} \int_0^\infty \langle \sigma_{\alpha\beta}(t) \cdot \sigma_{\alpha\beta}(0) \rangle dt \qquad (2)$$

Here, $k_B$ is the Boltzmann constant, $T$ is the temperature, $V$ is the volume of the simulation cell (based on $T$), and $\sigma_{\alpha\beta}$ is one of the components of the stress tensor. The ACF is averaged over three stress components being $\sigma_{xy}$, $\sigma_{xz}$, and $\sigma_{yz}$ and was evaluated using the 40 ps correlation length. The stress tensor data was obtained from the NNMD runs from over 7 ns of simulations for two temperatures at their equilibrium densities. Typically, in cases where significant network formation occurs, [6], [27], [39] a much longer trajectory is required for the stress ACF to converge to zero. However, in the case of $AlCl_3$, 7 ns of equilibrated trajectory was sufficient for the convergence of the ACF function as shown by the plateau in evaluated viscosities in **Figure S7**.

**RESULTS & DISCUSSION**

Our findings are organized in this section as follows. First, multiple XC functionals are evaluated to identify the most accurate method for the $AlCl_3$ system. The performance of each functional is based on how well they reproduce experimental Raman spectra, neutron structure factor, and liquid $AlCl_3$ density at 498 K. After benchmarking, the optimal DFT method is used to develop MLIPs with both the MLFF and NNIP architectures, which are then used to predict the melt structure and liquid densities. Next, using the two NNIPs, direct simulations of the two-phase systems are used to predict the liquid-vapor phase diagram of $AlCl_3$. By using both NNIPs, we demonstrate the importance of including low density configurations for accurate phase diagram predictions. Finally, the surface tension and melt viscosities of $AlCl_3$ are evaluated with both NNIPs and compared with the available experimental values [79].

**Determining appropriate DFT method for $AlCl_3$ molten salt**

To evaluate the appropriateness of the chosen DFT method to accurately represent the $AlCl_3$ system in the AIMD simulations, each functional in **Table 1** was evaluated by comparing three computed observables: Raman spectra, neutron structure factor (S(q)), and equilibrium density at 498 K and 4.33 bar.

*Raman spectra*

The Raman spectrum of $AlCl_3$ at 498 K was computed using all six XC functionals listed in **Table 1** from the last 60 ps of the equilibrated AIMD trajectories in the NVT ensemble at the experimental density. The computed spectra are compared against the experimental spectrum from ref. [30] in **Figure 1.** The wavenumbers and intensities of each computed spectrum were scaled to match the main peak location and height in the experimental spectrum at 340 $cm^{-1}$. As seen in **Figure 1**, all six functionals exhibit roughly the same peak locations as the experimental spectrum, showing good agreement with the experimental spectrum; however, some differences in the relative intensities at each peak location exist. SCAN and optB88-vdW align the best with experiment when considering the relative intensities between peaks located at 120 $cm^{-1}$ and 340 $cm^{-1}$. SCAN-RVV10 also aligns relatively well for the same peaks, while PBE-D3, SCAN, and optB86b-vdW exhibit slightly larger deviations. Importantly, none of the XC functionals' spectra suggest the presence of any species other than the molecular $Al_2Cl_6$ dimer, which is the expected structure in pure $AlCl_3$ melts.

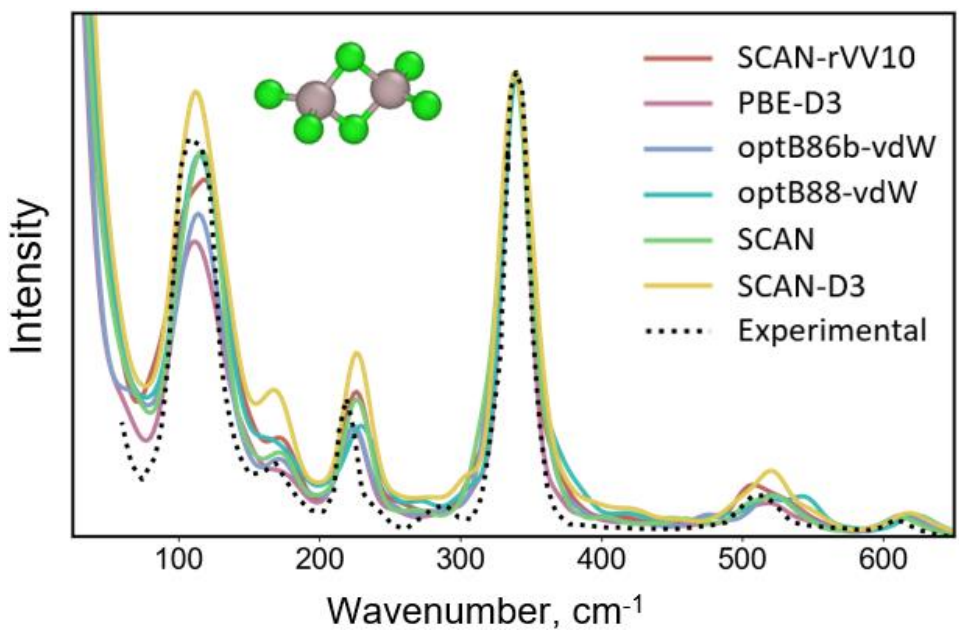

**Figure 1.** Experimental Raman spectra **[30]** compared with AIMD-computed spectra for $AlCl_3$ simulated at experimental density at 498 K. AIMD simulations were performed using SCAN-rVV10, PBE-D3, optB86b-vdW, optB88-vdW, SCAN, and SCAN-D3 density functionals. Inset shows the structure of the $Al_2Cl_6$ dimer.

*Neutron structure factors*

Using the same NVT AIMD trajectories as for the Raman spectra calculations, the total structure factors (S(q)) for all tested XC functionals were computed using ISAACS code [80] and was then compared against the experimental [32] data in **Figure 2**. Here, unlike the Raman spectra, we find very little differences between the computed and experimental structure factors. This suggests that each functional is capable of accurately describing the liquid structure of $AlCl_3$, which consists entirely of $Al_2Cl_6$ dimers. It should be noted that key structural correlations and features (described in our previous study on this system mixed with KCl [26]) are clearly identified in **Figure 2**. This includes the pre-peak feature (centered at $q \sim 1$ Å$^{-1}$), which represents two close-contact $Al_2Cl_6$ dimers, and the –Al–Cl–Al–Cl– charge alternation feature (centered at $q \sim 2.0$ Å$^{-1}$) [26].

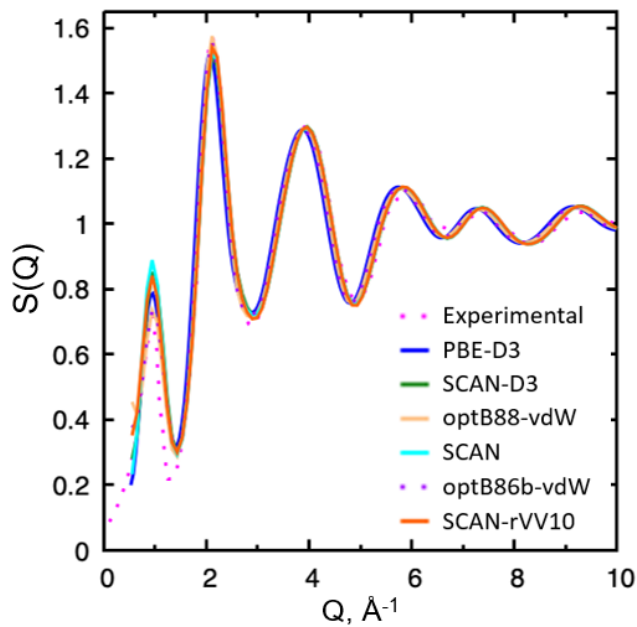

**Figure 2.** Comparison of the experimental neutron structure factors **[29]** with the computed ones for $AlCl_3$ simulated at experimental density at 498 K from the AIMD simulations performed using SCAN-rVV10, PBE-D3, optB86b-vdW, optB88-vdW, SCAN, and SCAN-D3 density functionals.

*Bulk Liquid Density at 498 K*

The equilibrium density of liquid $AlCl_3$ at 498 K was the final metric we used for benchmarking the performance of each XC functionals. Using each DFT method, the density was estimated by performing 100 ps of isothermal-isobaric AIMD (NPT) with the pressure and temperature set to 4.33 bar and 498 K, respectively. For all simulations, the initial density was set to the experimental density value of 1.21 g/cm$^3$. As the simulations proceeded, the systems densities for each functional (except SCAN) was evaluated in 10 ps block averages (displayed in **Figure 3**). The experimental AlCl$_3$ density at the same conditions is shown using the dashed horizontal line at 1.21 g/cm$^3$ for comparison. SCAN was omitted altogether as it significantly underpredicted the density and the corresponding AIMD simulations were not continued past 15 ps. For all other cases, the mean and standard deviation of the density for the last 5 blocks (50 ps) are shown in parentheses next to the XC functional name in the legend.

Among all the XC functionals evaluated, PBE-D3 demonstrated the best accuracy, underestimating the density by about 0.06 g/cm$^3$ (<5% error). Both SCAN-D3 and SCAN-RVV10 performed similarly, each underpredicting the density by ~0.2 g/cm$^3$ (~16% error). Surprisingly, despite their more sophisticated treatments of dispersion interactions, both optB86b-vdW and optB88-vdW were found to drastically overestimate the AlCl$_3$ density, yielding density values of 1.54 ± 0.05 g/cm$^3$ (27% error) and 1.82 ± 0.03 g/cm$^3$ (50% error), respectively. Despite being the least computationally demanding method, PBE-D3 exhibited the best agreement with the experimental density (together with a good match to the experimental Raman spectrum and structure factor), making it the best choice for modeling the AlCl$_3$ system. As a result, PBE-D3 was chosen to generate the training data for both MLIPs.

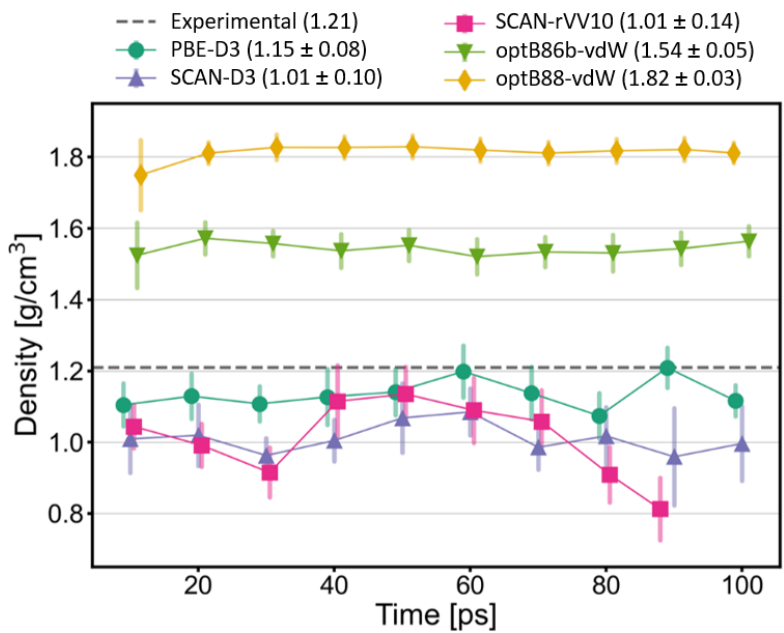

**Figure 3.** Density variations during AIMD for molten AlCl$_3$ at 498 K and 4.33 bar. Each point shows the mean density for 10 ps blocks and the error bars represent the standard deviation. The mean and standard deviation of the density (in g/cm$^3$) for the last 50 ps of each functional are shown in parentheses in the legend. SCAN was omitted because it rapidly dropped from the experimental value and was not run longer than 15 ps.

**Structural Analysis using Radial Distribution Functions**

The local coordination behavior of Al-Al, Al-Cl, and Cl-Cl is depicted using the partial radial distribution functions (RDFs). **Figure 4** shows and compares the Al-Al, Al-Cl, and Cl-Cl partial RDFs from NVT MD simulations at 498 K using PBE-D3, MLFF, NNIP-1, and NNIP-2. For all methods, a double peak is observed in the Al-Cl RDFs, which is attributed to the slight difference in Al-Cl distances for the chloride anions bridging two aluminum tetrahedra and the terminal chlorides (see **Figure 4(b)**). In addition, all methods yielded a close to zero minimum in the Al-Cl RDF plot, reflecting strong coordination driven by the high polarizability of the multivalent $Al^{3+}$ cation. This was also previously observed in the case of other molten salt systems containing multivalent cation species **[6], [27]**. The RDF peak heights and positions agree well across all methods, confirming the accuracy of the $AlCl_3$ structure generated in our MLFF and NNIP-based MD simulations relative to AIMD.

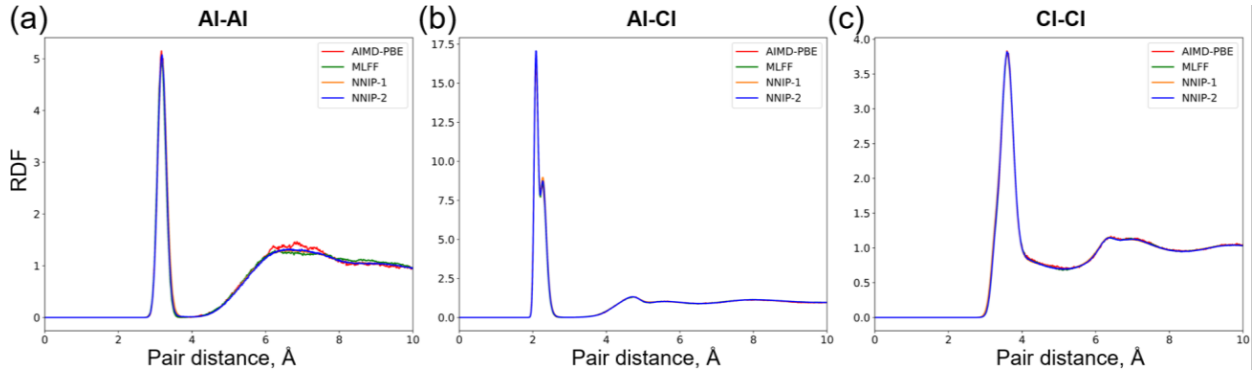

**Figure 4.** Comparison of (a) Al-Al, (b) Al-Cl, and (c) Cl-Cl radial distribution functions for $AlCl_3$ at T=498 K from PBE-D3, MLFF, NNIP-1, and NNIP-2 MD simulations.

**Bulk Liquid Densities as a Function of Temperature**

Upon validating the salt structure at 498 K, the developed MLIPs were employed to evaluate the liquid densities of $AlCl_3$ at wide range of temperatures (473-613 K). Specifically, PBE-D3 AIMD, MLFF MD, and NNMD simulations using both NNIPs were performed at 473, 498, 508, 543, 578, and 613 K. For their validation, the experimental values for equilibrium densities that are in equilibrium with its vapor at a given temperature were approximated using the empirical relation established in [4], computed using **Equation 3**, where, $b_0$ = 0.8279, $b_1$ = –9.02 × $10^{-4}$, $a_2$ = –181, $a_4$ = –546, and $T_c$ = 355.2 °C.

$$\rho(T\ [°C]) = b_0 + b_1 T + \sqrt{\frac{-a_2 - \sqrt{a_2^2 - 4a_4(T_c - T)}}{2a_4}}, \tag{3}$$

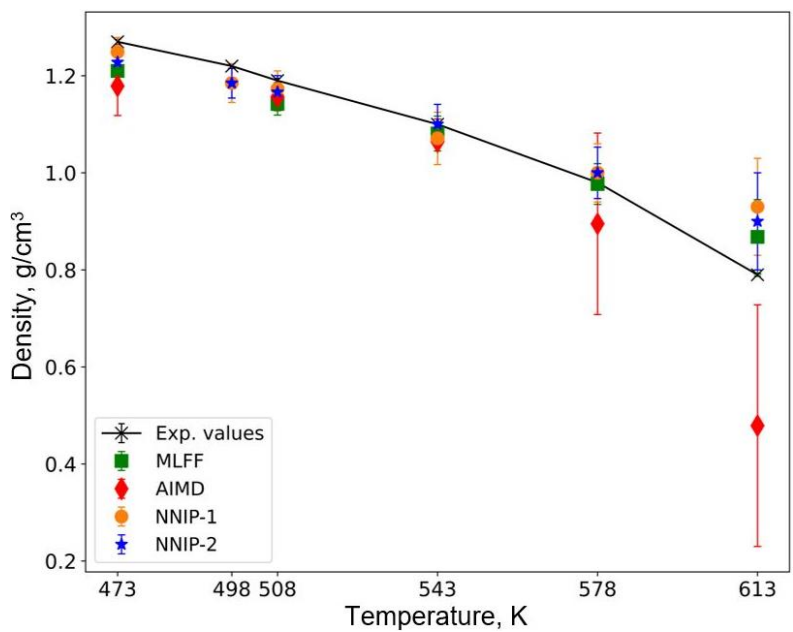

**Figure 5.** Comparison of predicted and experimental [4] liquid densities of $AlCl_3$ at temperatures of 473-613 K from AIMD PBE-D3, MLFF, NNIP-1, and NNIP-2 MD simulations.

**Figure 5** displays the $AlCl_3$ liquid densities from all MD simulations and compares them with the available experimental data **[4]**. The average densities over time from AIMD simulations are plotted in **Figure S8**, which highlights the large fluctuations in density, particularly at 578 and 613 K. Indeed, by performing block averaging on the time series of densities, we see that the standard error of the mean (SEM) density (inset in **Figure S8**) does not reach a plateau with increasing block sizes, which indicates insufficient sampling of the density at these higher temperatures of 578 and 613 K. As a result, the equilibrium density predictions either have considerable error bars (578 K) or are entirely unreliable (613 K). This limitation in density prediction is not necessarily surprising given the small system size, shorter timescales accessible to AIMD, and proximity to the critical temperature. In the case of MLIP-based MD simulations, due to eight times larger cell sizes and longer simulation times, the predicted density values are well-converged as well as the uncertainties associated with them are relatively low at all temperatures. Both MLIP architectures (MLFF and NNIP) slightly underpredict the densities at the lowest temperature, match more closely to the experimental values at intermediate temperatures, and slightly overpredict them at the highest temperature; however, in all cases, the experimental densities are still within the uncertainty of the predicted values. Importantly, each MLIP-predicted density is well converged and lies within the uncertainty of the AIMD-predicted values (except 613 K, where the AIMD results are unreliable, vide infra), which demonstrates the utility of simulating larger systems for longer times with DFT-trained MLIPs without a noticeable loss of accuracy. In all simulations, the uncertainty values for were obtained based on the 95% confidence interval of the mean estimate, while experimental uncertainties for liquid densities are reported to be 0.4% of their mean values **[4]**.

**Vapor-Liquid Two-Phase Equilibrium Simulations**

We evaluated the performance of both NNIP-1 and NNIP-2 potentials in predicting the liquid-vapor phase diagram of $AlCl_3$ by conducting two-phase equilibrium MD simulations at 473, 508, 543, 578, and 613 K for a system of 960 dimers. Further details of the simulations are provided in the Computational Methods section.

The snapshots from equilibrated two-phase simulations at each temperature are shown in **Figure 6(a-e)**. **Figure S5** shows the average density profiles for each snapshot (taken every 100 fs) as obtained from the equilibrated trajectories at various temperature. **Figure 6(f)** exemplifies the averaged density profile at 578 K from the direct simulation of the two-phase system over ~6 ns using NNIP-2. The uncertainty bars at each z-distance represent 95% confidence interval. The shifted density profiles from each 100 fs snapshot and the average density profiles for the other temperatures are shown in **Figure S5** and **Figure S6**, respectively. Using these average density profiles ($\rho_{average}$) obtained at each temperature, following Refs. **[16], [17]**, the vapor and liquid densities were determined by fitting each profile to the hyperbolic tangent expression given by **Equation 4**, where $d$ is the interfacial thickness, $z_0$ is the position of the interface, $\rho_l$ is the liquid density, and $\rho_v$ the vapor density.

$$\rho_{fit}(z) = \frac{\rho_l + \rho_v}{2} - \frac{\rho_l - \rho_v}{2} \tanh\left(\frac{z - z_0}{d}\right) \qquad (4)$$

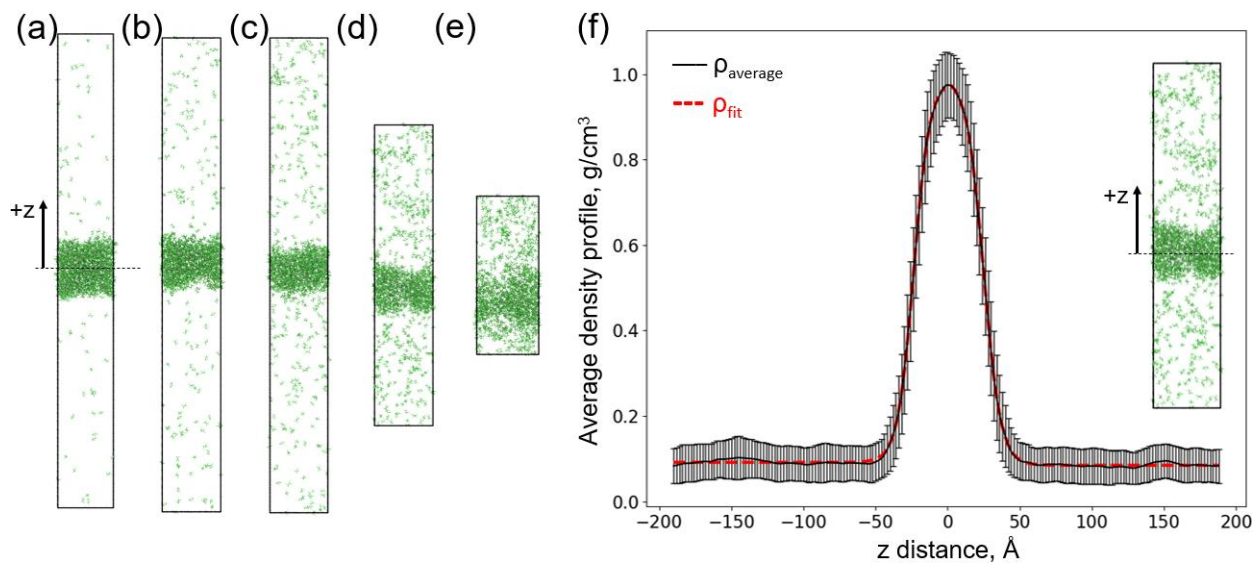

**Figure 6.** (a-e) Snapshots from equilibrated two-phase simulations at 473, 508, 543, 578, and 613 K, (f) Average density profile ($\rho_{average}$) obtained at 578 K using NNIP-2 as calculated from averaging density profiles obtained every 100 fs (shown in **Figure S5**) throughout the equilibrated trajectory. Here, the uncertainty bars at each z-distance represent 95% confidence intervals. The dashed red line is obtained from fitting $\rho_{average}$ profile using the hyperbolic tangent fit in **Equation 4**.
23

The average density profile ($\rho_{average}$) obtained from NNIP-2 at 578 K by fitting to the hyperbolic tangent function ($\rho_{fit}$) in **Equation 4** is illustrated in **Figure 6(f)** as a dashed red line. Fitted profiles for the rest of the temperatures are shown in **Figure S6**. From these fits, we extracted both liquid ($\rho_l$) and vapor ($\rho_v$) phase densities at each temperature, which are compared against the experimental data **[4]** in **Figure 7**. For both NNIPs, the uncertainty intervals for the predicted $\rho_l$ encompass the experimental values. Notably, as the simulation temperature increases, the NNIP-1-predicted values begin to deviate from experiment; whereas NNIP-2 shows much better agreement with experimental data for both $\rho_l$ and $\rho_v$. Although NNIP-1 agrees better with the experimental liquid density at the lowest temperature (T = 473 K), NNIP-2 is closer to the DFT liquid density (**Figure 5**), which is the more appropriate metric for comparison since an MLIP can only be as accurate as the reference DFT method on which it was trained. This improved agreement with the experimental vapor-liquid saturation curve from NNIP-2 can be attributed to the inclusion of lower-density cluster configurations (2-, 4-, and 7-dimers) in the NNIP-2 training set, as shown in the phase diagram in **Figure S1**. The addition of low-density configurations only had a relatively small influence on the liquid densities from NPT simulations (**Figure 5**), whereas a significant difference in liquid densities was seen in two-phase simulations (**Figure 7**) between NNIP-1 and NNIP-2 predictions. This indicates that these cluster configurations were critical for accurately describing the vapor phase, which consequently improved the liquid density for NNIP-2 over NNIP-1 in two-phase simulations.

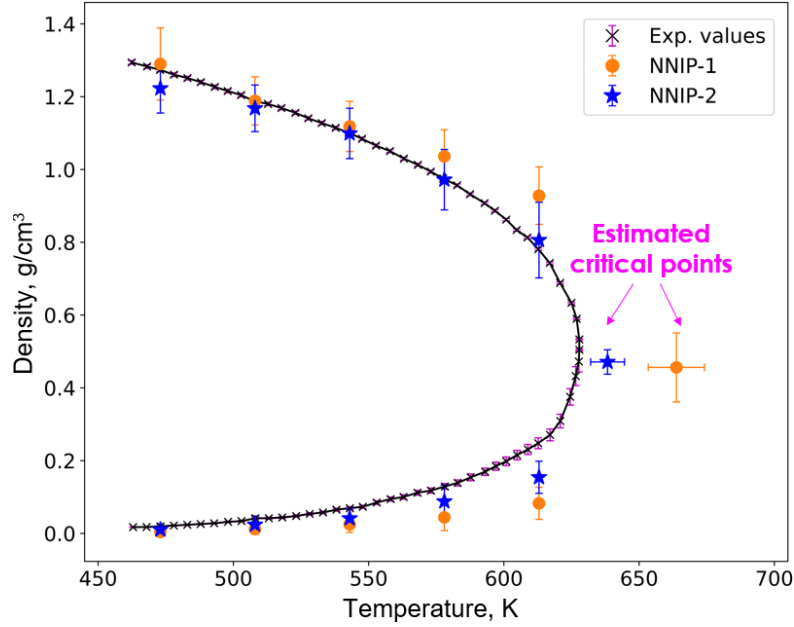

**Figure 7.** AlCl$_3$ vapor-liquid phase diagram obtained from two-phase NNMD simulations with NNIP-1 and NNIP-2 as compared to experimental data **[4]**.

Upon evaluating the liquid and vapor densities at five temperatures using both NNIPs, we determined their critical temperature and density values by simultaneously fitting the calculated, subcritical liquid and vapor densities to the scaling relations given in **Equation 5(a)** and **5(b)**. In these equations, the critical exponent $\beta$ for the three-dimensional Ising model is taken to be 0.326 **[81]**, while $A$ and $\Delta\rho_0$ are the system-specific parameters obtained in the fitting process.

$$\frac{\rho_l + \rho_v}{2} = \rho_c + A(T_c - T) \qquad (5(a))$$

$$\rho_l - \rho_v = \Delta\rho_0 \left(1 - \frac{T}{T_c}\right)^\beta \qquad (5(b))$$

**Figure 7** marks critical points based on liquid and vapor density predictions from both NNIP-1 and NNIP-2. The NNIP-1 estimates a critical temperature of 663 ± 10 K and a density of 0.456 ± 0.1 g/cm$^3$, resulting in the overestimation of the critical temperature by over 36 K ($T_{c,exp} =$

**627.7 K**). While large uncertainty is associated with NNIP-1 computed critical density, the average value is underestimated by more than 10% ($\rho_{c,exp} = \mathbf{0.51 \pm 0.03}$ g/cm$^3$). Conversely, NNIP-2 provides more accurate predictions with a critical temperature of 638.3 ± 6 K and a density of 0.47 ± 0.03 g/cm$^3$, which only deviate from experiment by 10 K and ~7%, respectively. This trend is also consistent with previous studies that witnessed improved predictions of the critical temperature and density for water after incorporating lower density configurations into the NNIP training **[15], [16]**. Overall, by showing the dramatic improvement in the prediction of liquid and vapor densities—as well as the critical temperature and density—of pure AlCl$_3$, we clearly demonstrate the importance of incorporating lower density cluster configurations into developed NNIPs when directly simulating two-phase systems.

It is instructive to compare the liquid densities obtained from two-phase simulations (**Figure 7**) with those obtained from the NPT simulations (**Figure 5**) using the experimental pressure at which two phases are at equilibrium. We found that the calculated liquid densities from NNIP-2 agreed reasonably well within their reported uncertainties (shown in **Figure S9**), except at the highest temperature, implying that the experimental pressures we imposed in our NPT simulations are sufficiently close to the predicted equilibrium pressures (comparison shown in **Figure S10**). However, the equilibrium pressure from the two-phase NVT simulations could not be determined accurately due to large fluctuations in the total system pressure, resulting in high uncertainties (**Figure S10** and **Table S1**).

**Surface Tension**

To evaluate whether the choice of NNIP affects the interfacial properties, we also computed the surface tension using both NNIP-1 and NNIP-2. **Equation 6** was employed to obtain the surface tension for the two-phase system at each temperature [82], [83], where $L_z$ is the simulation cell length in the normal or z direction, and $P_{zz}$, $P_{xx}$, and $P_{yy}$ are the pressures in z, x, and y directions, respectively.

$$\gamma = \frac{L_z}{2}\left[\langle P_{zz}\rangle - 0.5\left(\langle P_{xx}\rangle + \langle P_{yy}\rangle\right)\right] \tag{6}$$

The results at 473, 508, 543, 578, and 613 K are reported in **Figure 8**. As expected, both NNIPs predict almost linearly decreasing trend in the surface tension with increasing temperature, albeit with slightly different slopes. Notably, NNIP-2 predicts a statistically lower surface tension compared to NNIP-1 at the lowest temperature, which is consistent with the lower liquid density obtained for NNIP-2 at this temperature. No experimental data is available for comparison with the simulation results.

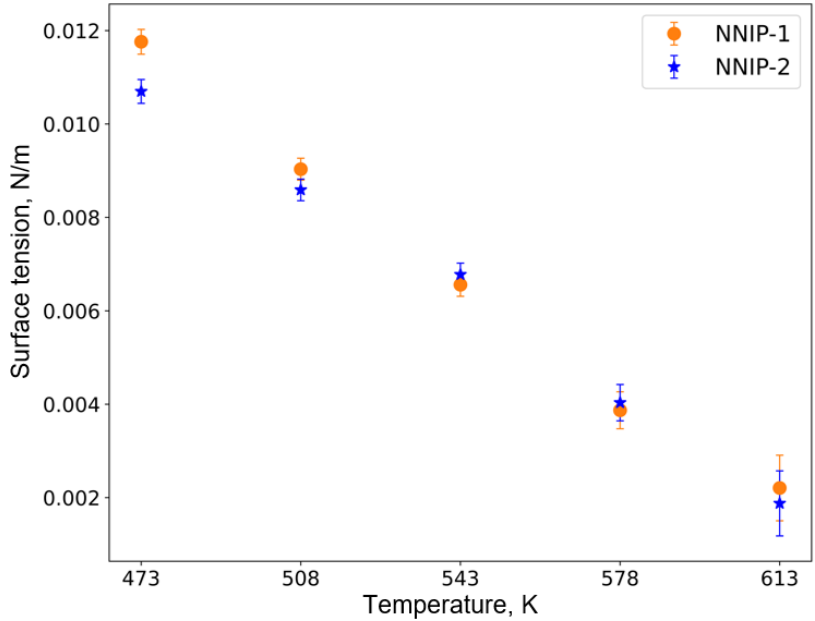

**Figure 8.** Comparison of the surface tension for AlCl$_3$ derived from two NNIP models. The uncertainties were calculated using 95% confidence intervals based on the normal and tangential pressure fluctuations at the equilibrated trajectory.

**Viscosity**

Viscosity is an important bulk property for predicting and optimizing the thermohydraulic performance of molten salts for various high temperature applications. We examined how the inclusion of low-density configurations—which is important for modeling the vapor-liquid phase diagram—influences bulk properties beyond liquid density, such as viscosity. The Green-Kubo method [78] was used to evaluate the viscosity of liquid AlCl$_3$ at 473 and 498 K (**Equation 2** in the Computational Methods Section). **Figure 9** presents a comparison of the NNMD-calculated and experimental viscosity values over the temperature range of 470–498 K [79]. Both NNIP models predict viscosity within 7% of the experimental values. Although the absolute deviations in viscosity values predicted by both NNIPs are similar, NNIP-2 more accurately captures the trend with temperature. The results indicates that the bulk properties of AlCl$_3$ are not negatively impacted by the inclusion of low-density configurations and that the NNIP-2 model performs much better under different thermodynamics conditions (single phase NPT and two-phase NVT simulations)

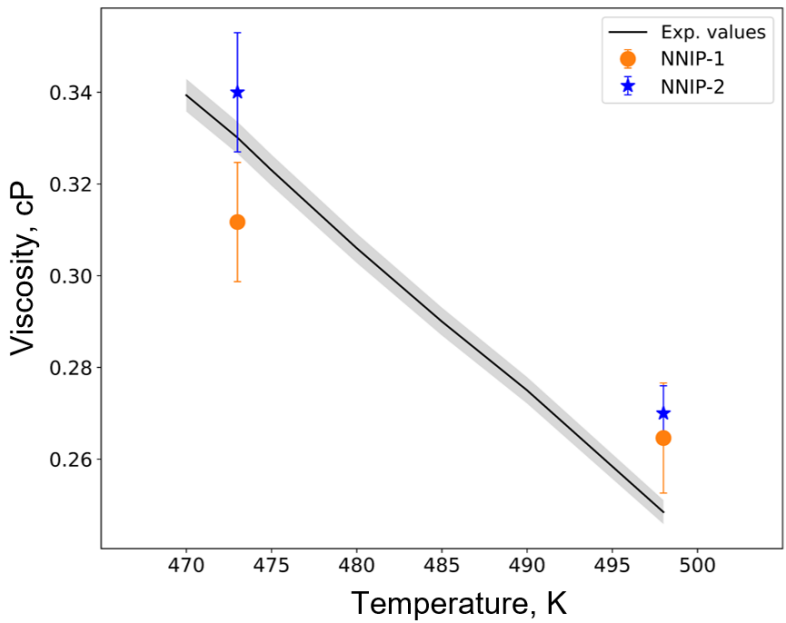

**Figure 9.** Comparison of viscosity for AlCl$_3$ from two NNIP models with the experimental results. The uncertainties from the NNIPs predictions are given by the standard deviations The reported uncertainty of the experimental data is 1.05% [79].

**CONCLUSION**

In this work, we demonstrated the feasibility of applying MLIPs developed based on DFT-based AIMD simulations to evaluate the vapor-liquid phase diagram for AlCl$_3$ through two-phase equilibrium simulations between 463 and 613 K. Two independent MLIP architectures, namely VASP-based MLFF and DeePMD-based NNIP, were developed using the PBE-D3 method, which was identified as the most suitable DFT method among several XC functionals and dispersion methods considered, including SCAN, SCAN-D3, SCAN-RVV10, optB86b-vdW, and optB88-vdW. PBE-D3 was chosen due to its superior performance in reproducing the experimental liquid phase density of AlCl$_3$, in addition to the good agreement with the experimental Raman spectrum and the neutron structure factor. Both MLIP architectures were used to evaluate the structure and liquid densities of $AlCl_3$ and validate the predictions against the AIMD simulations. By taking advantage of the scalability and efficiency of the developed NNIPs, direct two-phase MD simulations were performed to compute the liquid-vapor phase diagram of $AlCl_3$. We trained two NNIPs: NNIP-1, which is based exclusively on liquid configurations, and NNIP-2, which incorporates both liquid and low-density cluster configurations in the training. By including these additional, low-density configurations, the NNIP-2 model demonstrated superior agreement with the experimental critical temperature and density of $AlCl_3$, deviating by only 10 K (~3%) and 0.03 g/cm$^3$ (~7%) of the experimental data. This is a notable improvement over the NNIP-1 model, which yielded a critical temperature that was overpredicted by nearly 36 K (>10%) and a critical density that was underpredicted by 0.053 g/cm$^3$ (>10%). The NNIP potentials were further evaluated to predict the surface tension and viscosity for $AlCl_3$ at the temperatures between 473 and 613 K. While no experimental data are available for the surface tension, the trend of viscosity with temperature obtained from NNIP-2 shows better agreement with experiment. We therefore conclude that the NNIP training on both liquid- and gas-phase configurations is required to perform well across different thermodynamic conditions.

Overall, the demonstrated applicability of NNIPs shown in this work opens up the possibility of developing robust and reliable NNIPs to advance the understanding of complex neoteric liquids across a wide configurational phase space. In light of these findings, there is an urgent need to evaluate the vapor pressure in other complex binary and ternary molten salt systems relevant to nuclear energy applications. In contrast to the neutral $Al_2Cl_6$ comprising molten $AlCl_3$, typical molten salt compositions are ionic, making long-range electrostatic effects more significant for these systems. Addressing these effects may involve explicitly incorporating long-range electrostatic interactions into the MLIP architecture, an area currently under active research [84], [85], or employing a different computational formalism that entails computing chemical potentials in two phases without constructing an actual interface. Such research could facilitate computational screening of molten salt compositions for suitable operating temperature ranges, modeling accident scenarios, and supporting safety analysis.

**ASSOCIATED CONTENT**

**Supporting Information**.

The SI file contains 10 Figures and 1 Table containing additional information on training datasets, MLIP (MLFF and both NNIPs) testing and their comparison, analysis details for densities and viscosity calculation.

**AUTHOR INFORMATION**

**Corresponding Author**

*chahalr@ornl.gov, bryantsevv@ornl.gov*

**Author Contributions**

RC led and performed DeePMD simulations, NNIPs development, including subsequent computational analyses. LDG led and performed MLFF training and subsequent MD simulations and analysis. VSB designed and administered the research project, performed AIMD and Raman spectra simulations. SR contributed to the AIMD analysis and editing of the manuscript. The original manuscript draft was prepared by RC, LDG, and VSB. All authors proofread and approved the manuscript.

**Notes**

This manuscript has been authored in part by UT Battelle, LLC, under contract DE-AC05-00OR22725 with the US Department of Energy (DOE). The US government retains and the publisher, by accepting the article for publication, acknowledges that the US government retains a nonexclusive, paid-up, irrevocable, worldwide license to publish or reproduce the published form of this manuscript, or allow others to do so, for US government purposes. DOE will provide public access to these results of federally sponsored research in accordance with the DOE Public Access Plan (http://energy.gov/downloads/doepublic-access-plan).

**ACKNOWLEDGMENT**

The authors would like to thank the NERSC staff, especially Neil Mehta, for software support and assisting with a parallel installation of the DeePMD kit on Perlmutter. We are grateful to Rabi Khanal for his assistance in performing the structure functions calculations. This work was supported by the Office of Materials and Chemical Technologies within the Office of Nuclear Energy, U.S. Department of Energy. This research used resources of the Oak Ridge Leadership Computing Facility at the Oak Ridge National Laboratory, which is supported by the Office of Science of the U.S. Department of Energy under Contract No. DE-AC05-00OR22725. Additionally, this research used resources of the National Energy Research Scientific Computing Center (NERSC), a U.S. Department of Energy Office of Science User Facility located at Lawrence Berkeley National Laboratory, operated under Contract No. DE-AC02-05CH11231.

**For Table of Contents Only**

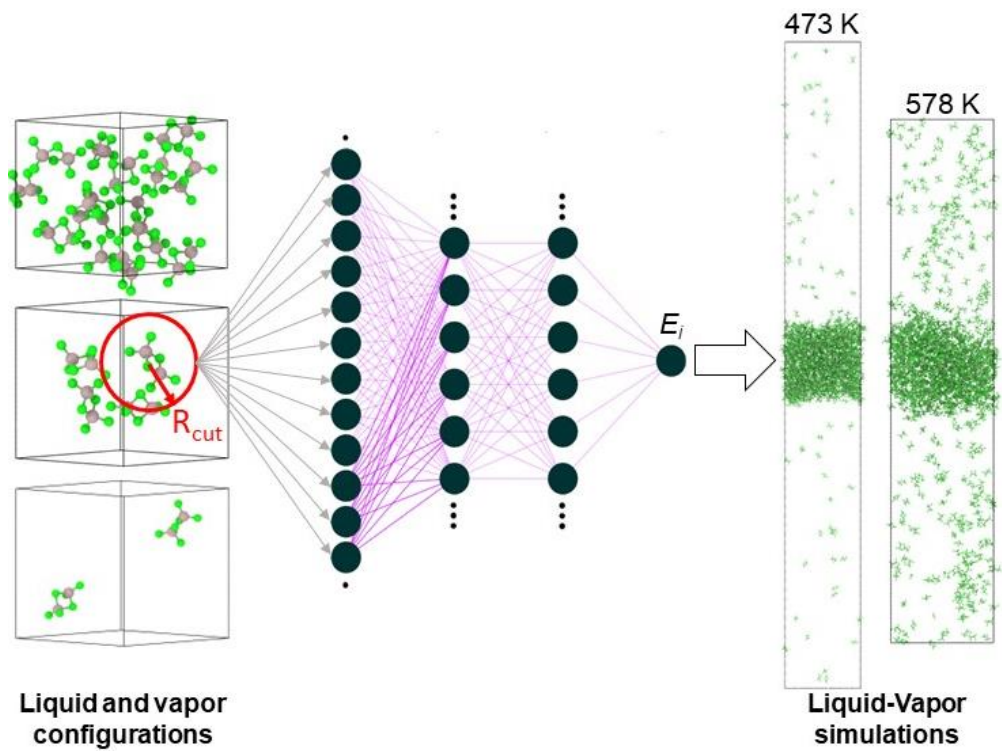

Liquid and vapor configurations — Liquid-Vapor simulations